# Ultra Low Specific Contact Resistivity in Metal-Graphene Junctions via Atomic Orbital Engineering

*Vikram Passi[1]‡, Amit Gahoi[2]‡, Enrique G. Marin[3]‡, Teresa Cusati[3], Alessandro Fortunelli[4], Giuseppe Iannaccone[3], Gianluca Fiori[3], Max C. Lemme[1,2]\**

[1] AMO GmbH, Advanced Microelectronic Center Aachen, Otto-Blumenthal-Str. 25, 52074 Aachen, Germany

[2] Chair of Electronic Devices, RWTH Aachen University, Otto-Blumenthal-Str. 2, 52074 Aachen, Germany

[3] Dipartimento di Ingegneria Dell'Informazione, Università di Pisa, 56122 – Pisa, Italy

[4] CNR-ICCOM, Consiglio Nazionale delle Ricerche, 56124 – Pisa, Italy

Abstract

A systematic investigation of graphene edge contacts is provided. Intentionally patterning monolayer graphene at the contact region creates well-defined edge contacts that lead to a 67% enhancement in current injection from a gold contact. Specific contact resistivity is reduced from 1372 Ωμm for a device with surface contacts to 456 Ωμm when contacts are patterned with holes. Electrostatic doping of the graphene further reduces contact resistivity from 519 Ωμm to 45 Ωμm, a substantial decrease of 91%. The experimental results are supported and understood via a multi-scale numerical model, based on density-functional-theory calculations and transport simulations. The data is analyzed with regards to the edge perimeter and hole-to-graphene ratio, which provides insights into optimized contact geometries. The current work thus indicates a reliable and reproducible approach for fabricating low resistance contacts in graphene devices. We provide a simple guideline for contact design that can be exploited to guide graphene and 2D material contact engineering.

The extraordinary electronic, optoelectronic and mechanical properties of graphene make it a promising candidate as a technology booster for micro- and nanoelectronics applications. Examples include radio frequency electronics,[1,2] integrated photodetectors,[3–5] and nanoelectromechanical systems.[6,7] One of the major bottlenecks limiting the performance of graphene-based devices is the large and varying value of specific contact resistivity ($R_C$) between metal contact electrodes and graphene.[8–11] When a metal is brought into contact with graphene, a junction with high contact resistivity is created, typically attributed to the low density of states (DOS) in graphene in particular when the Fermi level is near the Dirac point.[11] Although ab-initio calculations provide deeper insights into the contact problem, they also highlight the importance of the metal.[12–14] Experimentally, various methods have been reported to reduce $R_C$: one of the most common approaches is post-metallization annealing.[15–17] Other methods aim to modify the graphene prior to metallization in a random manner, such as low power oxygen plasma etch (with or without post-metallization annealing),[18] ozone pre-treatment,[19] intentional doping of graphene below the contact metal,[20] and ion beam irradiation.[21,22] A more deterministic approach is the formation of "edge"-contacts, where the graphene under the contact is partially removed by lithographic methods to enable the formation of covalent bonds between graphene and metal. This idea was proposed by means of an ingenious contact geometry by Wang [23]. Subsequently, the partial removal of the graphene under the contact by lithography, plasma or ion bombardment allowed a more versatile contact design. In particular, Smith *et al.*[24] investigated edge patterning of graphene with rectangular cuts under palladium (Pd) and copper (Cu) contacts with the transfer length method (TLM). The conclusion of this study was extended by Park *et al.*[20], who considered also titanium (Ti) contacts and analyzed the combined effect of molecular doping. Meersha *et al.*[25] investigated edge contacts to platinum (Pt) and Cu and discussed also different processing techniques for the creation of the covalent bonds. A different pattern consisting of round holes has been explored by Song et al.[26] in a compound gold/palladium (Au/Pd) metal contact. Round holes have also been investigated by Anzi et al.[27] for a large variety of metals including Au, Pd, silver (Ag), aluminum (Al), and nickel (Ni). The impact of the contact-edges on a low contact resistance, has been confirmed by Kelvin Probe Force Microscopy in Ni/Al contacts [28].

In this work, we investigate systematically the influence of size and density of round holes in graphene under Au contact metal, and the influence of the resulting hole-to-graphene ratio. In contrast to previous studies [23–27], where understanding of the trade-off between the beneficial role of the edge-contacts and the prejudicial impact of the etched graphene was limited to a qualitative perspective, we provide deeper insights into optimal contact configurations studying patterns of holes with varying diameter down to 50 nm. In addition, we analyze the impact of a varying charge carrier concentration in the graphene under the contact and rationalize the phenomenon of improvement in current injection associated with patterning by first-principles-based simulations. Finally, we provide engineering guidelines for reliable contact optimization.

All experiments have been carried out using chemical vapor deposited graphene that has been transferred onto silicon chips with 85 nm of thermal oxide and processed to obtain transmission line measurement (TLM) test structures.[23,27–29] Electron beam lithography and reactive ion etching have been employed to define holes with diameters ($h_D$) of 50 nm, 100 nm, 200 nm, 500 nm and 1 μm in the graphene at the contact areas prior to metal deposition. These make up a set of edge-contacted graphene devices. In addition, surface (top) contacts without holes have been fabricated. Details of the fabrication process are described in the methods section.

Schematics of surface and edge contacts are shown in Figure 1a-c. A scanning electron micrograph taken before and after graphene etching with lithographically patterned holes is shown in Figure 1d-e. The results reported in [24] have structures with a fixed $h_D$ of 200 nm albeit varying in number, however, in the present work not only variation of the hole diameter but also the position of the hole has been carried out together with simulation studies providing a deeper insight into the current injection mechanism for edge contact structures.

All devices have been measured in ambient conditions, i.e. at 21°C and relative humidity of 45%, using a Keithley SCS4200 semiconductor parameter analyzer. The TLM structures have been characterized as back-gated field effect transistors to record a set of transfer characteristics, i.e. drain currents $I_d$ as a function of back-gate voltages $V_{bg}$. One set of transfer $I_d$-$V_{bg}$ curves for varying hole diameters and a fixed channel length of 4 μm is shown in Figure 2a. More data from TLM structures with edge and with surface contacts is summarized in the supplementary information (Figure S3). From these graphs, TLM plots were extracted at different back-gate voltages, i.e. at the charge neutrality point (in our case at $V_{bg} \approx 38$ V) and at a large back-gate overdrive ($V_{bg}$ = -40 V) as shown in Figure 2b. This allowed extraction of the specific contact resistivity ($R_C$ x W) in each case with high fidelity; the R² value was greater than 0.984 in all cases. We found that for devices with hole diameters of 200 nm, 100 nm and 50 nm, the extracted $R_C$ values were lower for edge contacts than for surface contacts. However, we also observed that for devices with hole diameters of 500 nm and 1000 nm the extracted $R_C$ values were higher than for surface contacts.

Next, we analyzed the measured specific contact resistivity as a function of the contact perimeter under each contact ($C_P$), because this determines the total length of graphene edges available for each contact. The contact perimeter is defined as the product of one hole perimeter times the number of holes per contact. The unfilled red triangles and black squares are values calculated at back-gate voltage ($V_{BG}$) of ~38 V (charge neutrality point) and -40 V, respectively. The values of the respective hole diameters are given alongside the data points as can be seen in Figure 3a. The variation of $R_C$ as a function of the hole diameter is shown in Figure 3b with the values of the contact perimeter of the respective holes given alongside the data points. $R_C$ clearly decreases with contact perimeter. At the Dirac point, a 67% decrease from 1372 Ωμm for a device incorporating surface contacts to 456 Ωμm for a device with contacts patterned with holes of diameter 200 nm is achieved. Electrostatically doping the graphene by applying a back-gate bias of $V_{bg}$ = -40 V increases the DOS in graphene and generally reduces $R_C$. In addition, we observe a reduction of $R_C$ from 519 Ωμm for surface contacts to 45 Ωμm for edge contacts, a substantial decrease of 91% and among the lowest $R_C$ values reported. The extracted experimental $R_C$ values taken at the charge neutrality point (~38 V) and at $V_{bg}$ = -40 V for different hole diameters are summarized in

Table *1*. A closer look reveals that for nearly identical total contact perimeters of 120 µm, i.e. for 500 nm holes and for 50 nm holes, the contact resistivity varies considerably: small hole diameters provide distinctly lower $R_C$. We therefore conclude that the contact resistivity does not just depend on the edge perimeter, but also on the remaining graphene under the metal contact available for current conduction towards the device channel, as was also noticed in graphene patterned with rectangular cuts in [24]. This observation allows designing optimal conditions for current injection and thus low contact resistance in graphene devices and devices based on two-dimensional materials in general.

The current flow path at the graphene/metal contact is through the edge.[30] The parameter transfer length ($L_{TK}$) described by equation (1) in [30] is a property of the system and depends on the contact material. The transfer length values extracted using the model proposed by Shockley [29] are 0.9 µm and 1.7 µm for conventional contacts and edge contacts (hole everywhere under the metal configuration as shown in Figure 1c) with hole diameters of 200 nm. The value of $L_{TK}$ for $h_D$ of 50 nm, 500 nm and 1000 nm are 1.2 µm, 1.5 µm and 0.8 µm; respectively. From these results, it is evident that $L_{TK}$ increases in the case of edge contacts compared to conventional contacts. For structures with $h_D$ =200 nm, 36.9% of the total contact length is responsible for current transmission compared to 19.5% for structures with conventional contacts. For conventional contacts with nickel as metal electrode, similar values of transfer length ($L_{TK}$ ~1 µm) have been reported in.[30] An increase in $L_{TK}$ values has also been reported in [26] for structures with graphene antidot arrays under a metal electrode to introduce edge contacts to graphene.

The theoretical study of the graphene-metal contact has been subject of intense study from ab-initio to semi-analytical models [31–34]. Here a multi-scale simulation approach ranging from first-principle calculations to current transport simulations has been adopted to provide physical insights into the main mechanisms at play in determining the contact resistivity and providing design rules to further reduce the contact resistivity. A schematic depiction of the simulated contact is shown in Figure 3c. The proposed approach is particularized for the experimentally fabricated structures, but the assumptions made do not hinder their application to other patterned structures as those e.g. in.[23] given that a dedicated analysis of the graphene-metal junction and the patterns are performed. The continuity equation (eqn. 1) has been solved to model the contact resistivity in the graphene layer

$$\nabla \cdot J_{2D}(y,z) = |J_{inj}(y,z)|, \qquad (1)$$

where, as shown in Figure 3c, $J_{2D}$ is the in-plane current density (A/m) given by $J_{2D} = \nabla + D\nabla$ , and $J_{inj}$ is the gold-to-graphene vertically injected current (A/m²), where , and $D$ are the mobility, density, and diffusivity of the carriers, respectively and  is the electrostatic potential. Assuming quasi homogeneous density, the diffusive term can be neglected, resulting in $J_{2D} \approx \nabla$ . The vertical injected current from the overlapping metal can be written to first order in the voltage drop ($V_m - $ ) between metal and graphene as $J_{inj} = G(V_m - )$ where, G is the gold to graphene conductance (S/m²) and $V_m$ is the metal electrostatic potential. Including these terms in equation 1, we obtain

$$\nabla \quad \nabla \ (y,z) = G\,(V_m - \ (y,z)). \qquad (2)$$

The regions of distinct nature have to be considered when solving equation 2 applied to the metal-graphene edge contact and are labeled in the sketch of Figure 3c. Here, the sigma bonds formed between the metal and the carbon atoms at the edge [35] result in a stronger chemical binding, leading to higher transmission of carriers compared to those derived from van der Waals bonds between the metals and the atoms at the graphene surface. To this purpose, we have considered two different sets of parameters (µ, ρ, G) for the graphene flake either close to or far from the edges. Density Functional Theory (DFT) calculations using the Quantum Espresso package[36] (see Supplementary Note 5) have been performed to provide an estimation of these parameters. The Au-graphene structure depicted in Figure 3d has been considered. A structural optimization has been performed, leading to equilibrium inter-atomic distances between the metal and the graphene edges of d = 1.9 Å, while it is d = 3.1 Å in the center (e.g. for the entire surface contacts). This edge distance has been employed to determine the edge conductance using a tight-binding Hamiltonian and the Landauer's formalism (see Supplementary Note 5), resulting in an edge conductance of $G_b = 9.8$ mS/µm. The deposition of metal leads to a shift of the energy at the Dirac Point ($E_d$) in graphene with respect to the Fermi level ($E_f$) of the system (see Figure 3c)[14] which has been extracted from DFT calculations. The calculated shifts $E_d - E_f$ are: 0.35 eV and 0.14 eV at the edges and the surface graphene, respectively, which translates into in carrier concentrations of $9.16 \times 10^{12}$ cm$^2$ and $1.6 \times 10^{12}$ cm$^2$, respectively. Given the size of the holes, we assume that the metal fills the etched regions as shown in Figure 3c. The metal-filled regions then behave as equipotential surfaces and can be modeled imposing Dirichlet boundary conditions for the solution of the differential equation [36] (see Supplementary Note 3). A potential reference at the right edge of the contact (indicated as $y_{max}$ in Figure 3c) has been imposed to estimate the contact resistivity.

The value of and the current density component along the transport direction (y), given by $J_{2D,y} = q\ n\ d\ /dy$, are obtained by solving equation 2; and eventually the contact resistivity $R_c$ in can be expressed as

$$R_c = V_m / \int\int J_{2D,y}(y,z)\big|_{y=ymax} dz, \qquad (3)$$

where the integral accounts for the total current flowing at the right contact end, i.e., $y = y_{max}$, indicated by the black line in Figure 3c.

The plot shown in Figure 3a-b compares experimental and simulated $R_C$ as a function of the $C_P$ for various hole diameters and $R_C$ as a function of the $h_D$ for various contact perimeter, respectively. The unfilled green circles are the values obtained by simulation. In the numerical model, square holes have been considered for the sake of simplicity, but the same conclusions hold also for round holes. Further details about the numerical solution of equation 2 are described in the Supplementary Note 5. The results of the numerical model are in very good agreement (within the experimental error interval) for all configurations except for the 500 nm-hole patterning, where the theoretical model predicts a ~20% lower contact resistivity than the one obtained experimentally.

The calculated data support the findings of our experiments (and others[24]): (1) patterning the contact plays an important role in $R_C$ reduction, including the choice of the hole size and distribution, (2) this role cannot be interpreted uniquely in terms of the total contact perimeter, since large variations of $R_C$ are observed for similar $C_P$ in the experiments and in the model. In particular, $R_C$ in the $h_D = 500$ nm case is almost twice as high as in the $h_D = 50$ nm configuration (1354 Ωμm and 620 Ωμm, respectively), even though it has very similar contact perimeter ($C_P = 118$ μm and $C_P = 117$ μm, respectively). A possible explanation is the prejudicial reduction of the remaining graphene area ($A_G$ is equal to 46 μm² and 58.6 μm², respectively), which negatively affects the graphene in-plane current transport. A summary of the experimental and modelled $R_C$ values for the different hole diameters, the corresponding contact perimeter and the $A_G$ area is shown in

Table 2.

Additional simulations have been performed on differently patterned contacts to shed light on the role of the total perimeter and the hole size. In particular, we have considered holes with sizes ranging from $h_D$ = 25 nm up to $h_D$ = 800 nm arranged in two columns at the contact edges. The contact dimensions are 12 μm x 5 μm, as in the experimental configurations. A different total number of holes have been considered depending on the hole size: 160 for $h_D$ =25 nm, 80 for $h_D$ =50 nm- $h_D$ =200 nm, 60 for $h_D$ =300 nm, 40 for $h_D$ =400 nm and $h_D$ =500 nm, 30 for $h_D$ =600 nm and 20 for $h_D$ =700 nm and $h_D$=800 nm. The same set of parameters obtained from first-principle calculations (and used to fit the experimental data) have been employed. The contact resistivity as a function of the total contact perimeter for the different investigated patterns is shown in Figure 4a. The remaining graphene area is shown aside to each value of $R_c$. We observe two trends in Figure 4a. For large contact perimeters (50 μm < $C_P$ < 80 μm) there is a large scatter in the $R_c$ values for the different $h_D$. The observed differences in $R_c$ are in good correspondence with the remaining graphene area (with the exception of d = 700 nm), which indicates that this latter factor is dominant over $C_P$. For smaller contact perimeters, $C_P$ < 40 μm, $R_c$ varies only slightly (for these hole sizes, $A_G$ > 95% of the contact area). In this representation, the simulated date exhibits considerable dispersion for similar values of $C_P$. It is therefore difficult to infer design guidelines from this format. Figure 4b plots $R_c$ as a function of the hole size. Here, a clear decreasing trend in the contact resistivity is observed when $h_D$ is reduced. The advantage of small holes might be understood in terms of its larger perimeter/area ratio as can be seen in Figure 4c. This ratio can be taken as a good measure of the trade-off between the beneficial and detrimental effects of patterning: the edges of the holes contribute to reduce $R_c$ thanks to its increased conductance, but the total hole area, where the graphene has been etched, deteriorates in-plane transport.[24]

Additional insight on the influence of the hole size on the contact resistivity is obtained from Figure 4d, where the current density $J_{2D,y}$ escaping through the channel-end (shown as red dashed line in Figure S5e) of the contact is plotted for several hole-size patterns (in particular at y = 4.8 µm, close to the position of the holes at one border). The current density oscillates between maxima corresponding to the position of the holes and minima associated to the inter-hole regions. As expected from Figure 4b, there is an increase in the average value of the current when the hole size is decreased. In addition, for larger hole sizes the minima-maxima excursion to average current density is big (71% in the 800 nm holes and 36% in the 600 nm hole) as compared to the smaller hole sizes (12% in the 100 nm holes and 0.7% in the 25 nm hole). The simulations reveal that a denser grid of smaller holes results in a graphene electrostatic potential closer to that of the metal (see Supplementary Note 6), and consequently in a lower metal-graphene resistivity. In addition, we have observed that $R_c$ is mostly determined by the patterning closest to the channel-end. A similar effect of current crowding at the edge of a metal-graphene contact has already described in the literature for non-patterned contacts.[8] Our results suggest a simple design rule for contact engineering: to reduce contact resistance, efforts should be directed towards patterning the region close to the contact end, and focus should be put on achieving patterns where the etched regions of graphene maximizes the perimeter to area ratio.

A substantial reduction in metal-graphene contact resistivity is achieved through patterning of holes in the graphene in the contact regions. This decrease amounts to 67% in unbiased graphene, where $R_C$ reduced from 1372 Ωµm for devices with surface contacts to 456 Ωµm for devices with patterned contacts. The experimental data shows a further reduction of $R_C$ when the graphene is electrostatically doped, resulting in a decrease from 519 Ωμm to 45 Ωμm, among the lowest values of contact resistivity reported so far. Multi-scale simulations quantitatively support and rationalize the present results and provide further insight in design rules for patterned contact engineering, showing the relevance of maximizing the perimeter-to-area ratio of the etched regions, and the relevance of the patterning close to the front end of the contact. The combined experimental and simulation data lead us to propose general guidelines for designing optimized contacts to graphene.

**Methods**

Experiment: Monolayer graphene has been grown on copper foil in a NanoCVD (Moorfield, UK) rapid thermal annealing tool using the chemical vapor deposition method described in [37]. Subsequently, a thin layer of poly-methyl-methacrylate (PMMA) has been spin coated onto the copper foil and baked at 180°C resulting in a thickness of 200 nm (measured by ellipsometry on a reference silicon-oxide substrate). Using an electro-chemical delamination method, graphene has been detached from the copper foil. The detached polymer supporting the graphene layer has been rinsed with water and transferred onto a pre-cleaned, thermally oxidized p-type silicon substrate with an oxide thickness of 85 nm. Structures comprising of the channel region and the holes have been defined in a single lithography step using hydrogen-silsesquioxane (HSQ) a negative tone electron beam resist, used directly as an etch mask for pattern transfer. Electron beam exposure has been carried out using a Raith EPBG-5000Plus electron-beam system at 100 keV. The unexposed resist has been developed using tetra-methyl-ammonium-hydroxide-25% solution and subsequently the unprotected graphene has been etched by oxygen plasma, resulting in the formation of well-defined channels and holes. Bilayer PMMA resist has been spin coated followed by electron beam lithography to expose windows in the resist. The exposed resist has been developed for 1 min using a mixture of MIBK and IPA solution. At this stage, the contact areas have been exposed in the resist and a short oxide etch using wet chemistry has been done to remove the HSQ from the graphene surface. Then 300 nm of gold has been deposited to form metal contacts to graphene and the excessive metal is removed by lift-off in suitable solvent. A thin layer of gold has been deposited at the backside of the substrate to act as the back-gate. A process flow with the main process steps is presented in the supporting information.

Density Functional Theory: First principles calculations have been performed using the Quantum Espresso package,[36] a plane wave basis set, a gradient-corrected exchange correlation functional (Perdew-Burke-Ernzerhof, PBE)[38] and ultrasoft pseudopotentials (US-PPs)[39] in scalar relativistic form.

The simulated Au-graphene structure can be seen in Figure 3c. The simulated configuration includes three continuous layers of Au and an interrupted layer of graphene, interacting with an interrupted layer of Au placed in the same plane. An orthorhombic cell has been used adjusting the graphene lattice size to that of Au. A zigzag orientation of the graphene edge has been chosen, because it is energetically more stable.[40] The x-direction includes the vacuum region, of 25 Å, in order to minimize the interaction between adjacent image cells. The Brillouin zone is sampled using $k_x=k_y=1$ and $k_z=8$. Dipole correction and the dispersion effects (Van der Waals corrections[41]) were included in the simulations. An energy cutoff of 40 Ry has been used for selection of the plane-wave basis set for describing the wave function and 400 Ry for describing the electron density cutoff. The geometry of graphene and gold was optimized keeping the upper two gold layers fixed and relaxing the position of the other gold and carbon atoms and the dimension of the cell. The curve corresponding to the Dirac point energy on the graphene ribbon in the interacting system was determined following the method used in reference[14], and shifting the Dirac point of the graphene ribbon underneath the metal with respect to a monolayer graphene by comparing the DOS projected on each carbon atom.

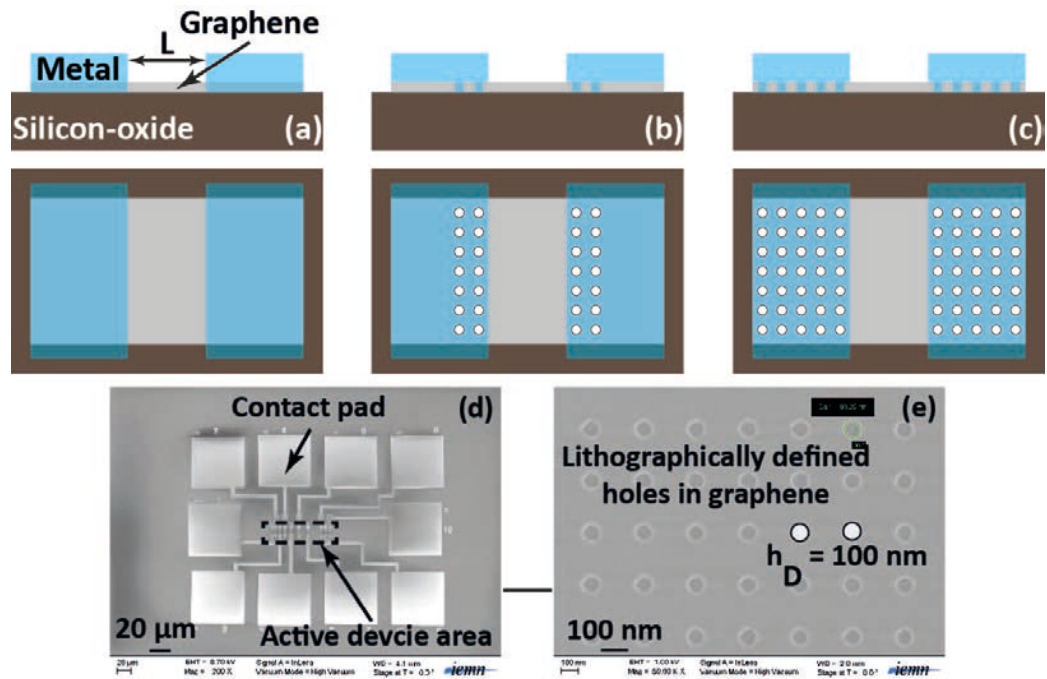

Figure 1: Schematic of the different contacts considered in this work. (a) Conventional contacts, (b) edge-contacts with holes in graphene positioned only under the metal edge, (c) edge-contacts with holes everywhere under the metal. (d) Scanning electron micrographs of entire device after fabrication and (e) after etching the unprotected graphene. The hole diameter is 100 nm.

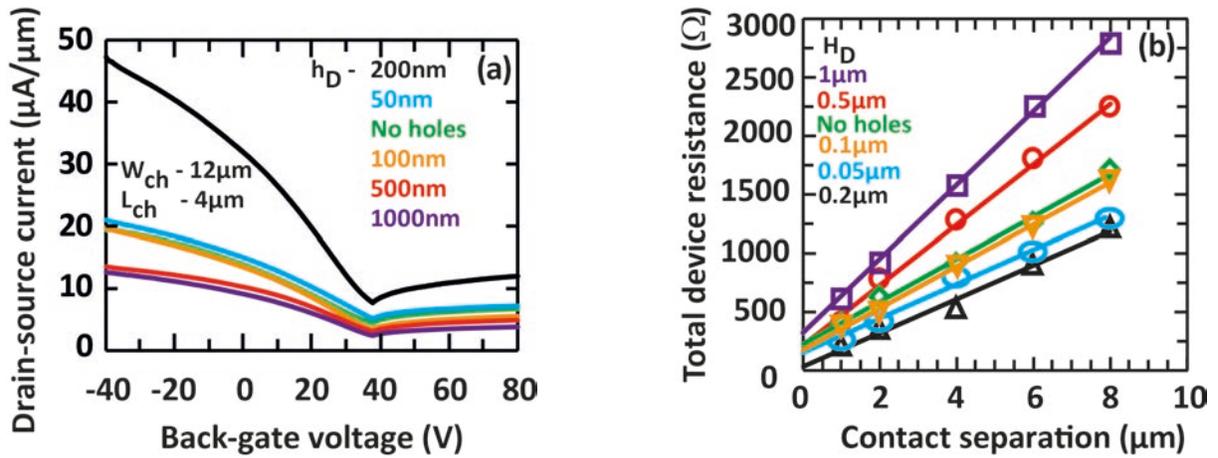

Figure 2: (a) Comparison of the measured drain current as a function of the applied back-gate bias, with surface and patterned contacts. The values of the hole diameter range from 50 nm up to 1 μm. The channel length is 4 μm and the channel width is 12 μm. (b) Transmission line measurements of test structures with different contact properties. The total resistance is normalized to the test structure (i.e. graphene) width to extract specific contact resistivity.

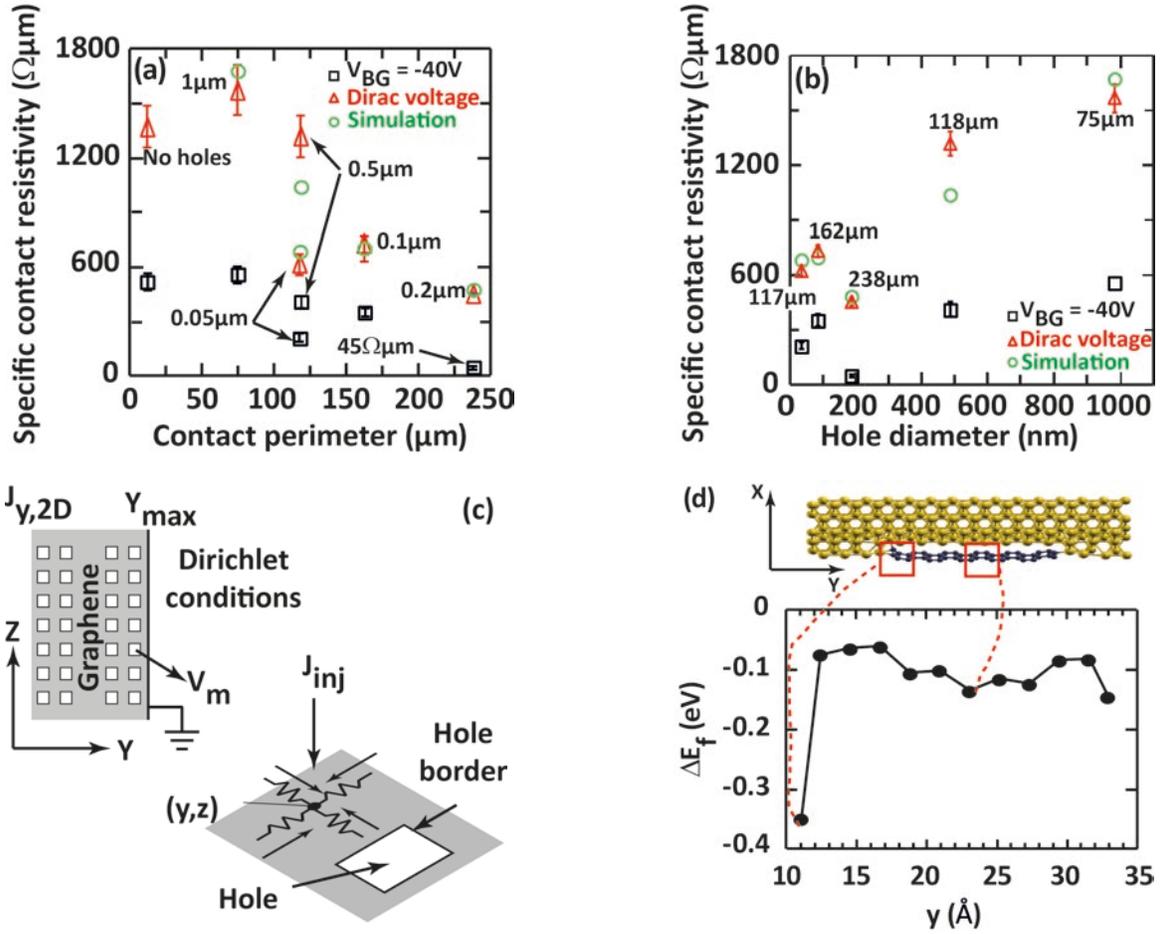

*Figure 3: Variation of the specific contact resistivity as a function of the (a) total contact perimeter and (b) hole diameter. The unfilled red and black symbols denote experimental values calculated at Dirac voltage and at an applied back-gate bias of -40V, respectively. The unfilled green circles denote numerical simulations. The experimental contact resistivity values were calculated at the Dirac voltage (between 37.8 V and 38.2 V) and the estimated error is ~10%. The adjacent labels in (a) indicate the diameter of the holes and in (b) indicate the contact perimeter values, respectively. (c) sketch of the structure considered in the numerical model, showing patterned graphene and the in-plane and vertically injected currents and (d) Sketch of the structure (gold-graphene) used for DFT simulation, and the Fermi energy shift with respect to the Dirac point obtained from the atomistic electrostatic potential analysis on carbon atoms.*

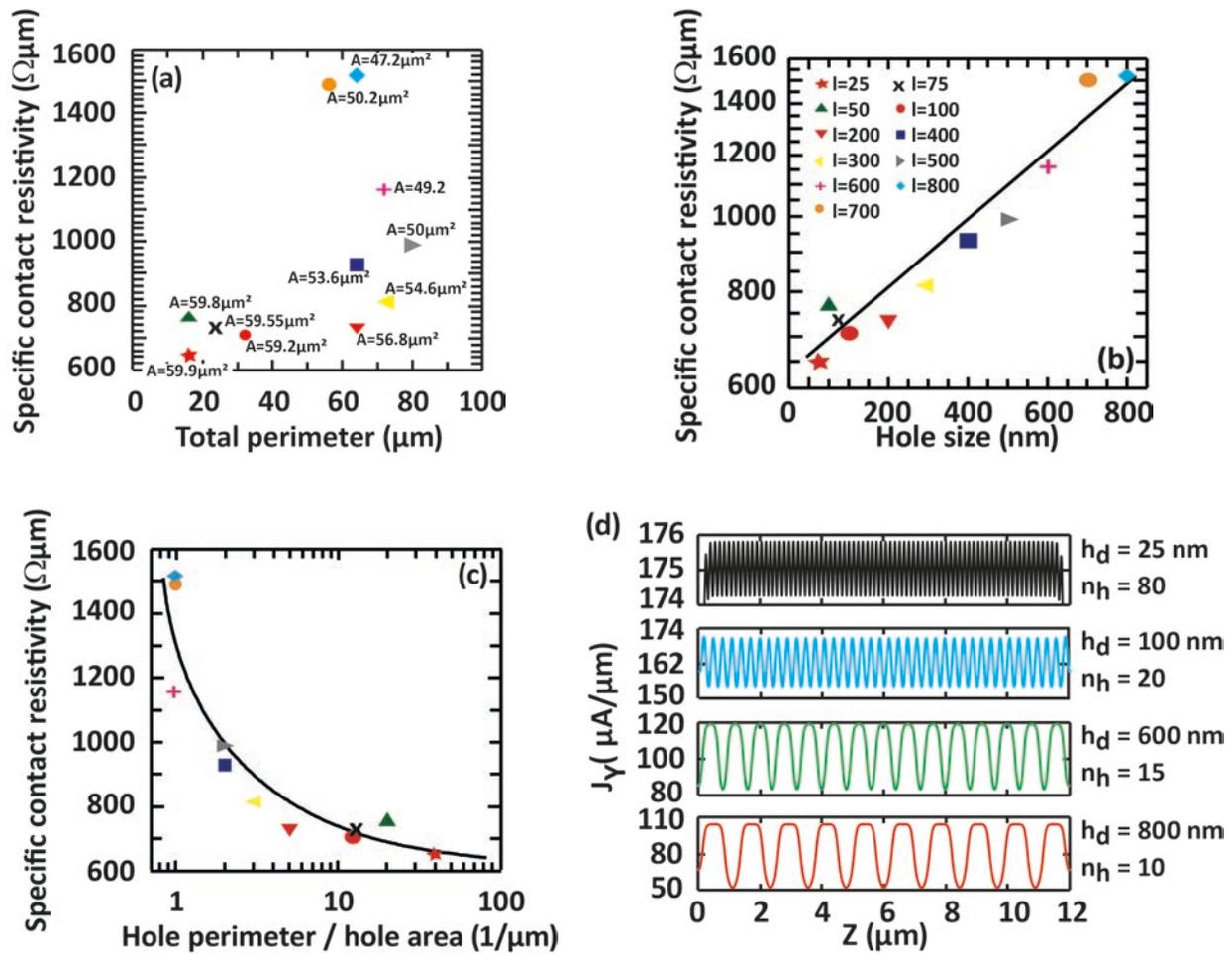

Figure 4: Simulated contact resistivity as a function of the a) total perimeter, b) the hole size, and c) hole perimeter/area ratio. In b) and c) the lines are a guide for the eyes. d) Current density through the channel-end of the contact, y = 4.8 μm, for several hole patterns.

Table 1: Summary of the hole diameter, contact perimeter, and the experimental value of contact resistivity extracted at the Dirac voltage (~38.2 V) and at applied $V_{BG}$ of -40 V, respectively.

| $h_D$ (nm) | $C_P$ (μm) | Experimental value of contact resistivity at Dirac voltage (Ωμm) | Experimental value of contact resistivity at $V_{BG}$=-40V (Ωμm) |
|---|---|---|---|
| **No holes** | 12 | 1372 | 519 |
| **50** | 117 | 620 | 212 |
| **100** | 162 | 732 | 352 |
| **200** | 238 | 456 | 45 |
| **500** | 118 | 1354 | 410 |
| **1000** | 74 | 1590 | 560 |

Table 2: Summary of the hole diameter ($h_D$), contact perimeter ($C_P$), remaining graphene area ($A_G$), and experimental and simulated contact resistivity values ($R_C$).

| $h_D$ (nm) | $C_P$ (µm) | $A_G$ (µm²) | Experimental $R_C$ (Ωµm) | Simulated $R_C$ (Ωµm) |
|---|---|---|---|---|
| **50** | 117 | 58.6 | 620 | 684 |
| **100** | 162 | 56.2 | 732 | 701 |
| **200** | 238 | 48.9 | 456 | 480 |
| **500** | 118 | 46 | 1354 | 1040 |
| **1000** | 74 | 44.3 | 1590 | 1681 |

AUTHOR INFORMATION

**Corresponding Author**

*lemme@amo.de, max.lemme@eld.rwth-aachen.de, Tel: +492418867200

**Author Contributions**

The manuscript was written through contributions of all authors. All authors have given approval to the final version of the manuscript. ‡These authors contributed equally.

ACKNOWLEDGMENT

The authors gratefully acknowledge funding from the European Commission through the European Regional Development Fund (EFRE-0801002, HEA2D), the European Research Council (ERC, InteGraDe, 3017311) and the Graphene Flagship (785219).